\documentclass{elsart3p}

\usepackage{epsfig}

\usepackage{amssymb}

\usepackage{xspace}

\def\cref#1{Chapt.\,\ref{#1}}
\def\Cref#1{Chapter~\ref{#1}}
\def\sref#1{Sect.\,\ref{#1}}

\def\fref#1{Fig.\,\ref{#1}}

\def\Fref#1{Figure~\ref{#1}}

\def\eref#1{(\ref{#1})}

\def\rref#1{Ref.\,\cite{#1}}

\def\rright{\textit{right}}
\def\LLeft{\textit{Left}}

\def\1{\footnotemark[1]}
\def\and{\& }
\def\Cerenkov{\v{C}erenkov\xspace}

\def\gcm2{g/cm$^2$\xspace}

\def\lg{{\rm lg}\xspace}

\def\modell{poly-gonato model\xspace}

\def\Xmax{$X_{max}$\xspace}

\def\Section#1{\section{#1}}

\begin{document}

\begin{frontmatter}



\title{The origin of galactic cosmic rays\thanksref{1}}
\thanks[1]{Invited talk given at the Roma International Conference on
           Astro-Particle physics (RICAP07) June 20th - 22nd, 2007.}


\author{J\"org R. H\"orandel}
\ead[url]{hoerandel.com}
\address{Radboud University Nijmegen, Department of Astrophysics, 
         P.O. Box 9010, 6500 GL Nijmegen, The Netherlands }

\begin{abstract}
 The origin of galactic cosmic rays is one of the most interesting unsolved
 problems in astroparticle physics. Experimentally, the problem is attacked by
 a multi-disciplinary effort, namely by direct measurements of cosmic rays
 above the atmosphere, by air shower observations, and by the detection of TeV
 $\gamma$ rays.  Recent experimental results are presented and their
 implications on the contemporary understanding of the origin of galactic
 cosmic rays are discussed.
\end{abstract}

\journal{Nuclear Instruments and Methods A}
\end{frontmatter}

\Section{Introduction}

The Earth is permanently exposed to a vast flux of highly energetic particles,
fully ionized atomic nuclei from outer space. The extraterrestrial origin of
these particles has been demonstrated by V. Hess in 1912 \cite{hess} and he
named the particles "H\"ohenstrahlung" or "Ultrastrahlung". In 1925 R. Millikan
coined the term "Cosmic Rays". They have a threefold origin.  Particles with
energies below 100~MeV originate from the Sun.  Cosmic rays in narrower sense
are particles with energies from the 100~MeV domain up to energies beyond
$10^{20}$~eV. Up to several 10~GeV the flux of the particles observed is
modulated by the 11-year cycle of the heliospheric magnetic fields.  Particles
with energies below $10^{17}$ to $10^{18}$~eV are usually considered to be of
galactic origin. A proton with an energy of $10^{18}$~eV has a Larmor radius
$r_L=360$~pc in the galactic magnetic field ($B\approx3$~$\mu$G). This radius
is comparable to the thickness of the galactic disc and illustrates that
particles at the highest energies can not be magnetically bound to the Galaxy.
Hence, they are considered of extragalactic origin.

The energy density can be inferred from the measured differential energy
spectrum $dN/dE$ \cite{halzenrhoe}
\begin{equation}
 \rho_E=\frac{4\pi}{c}\int \frac{E}{\beta} \frac{dN}{dE} dE ,
\end{equation}
where $\beta c$ is the velocity of particles with energy $E$.  For galactic
cosmic rays the major contribution to the total energy density originates from
particles with energies around 1~GeV. Such particles are strongly influenced by
the heliospheric magnetic fields. Outside the heliosphere, in the local
interstellar environment an energy density $\rho_E^{LIS}=1.1$~eV/cm$^3$ is
obtained (\rref{gaisserbuch} p.\,12).  The parameterization of the measured
galactic cosmic-ray flux according to the \modell \cite{pg} results in a
density $\rho_E^{gal}=0.43$~eV/cm$^3$. This is the measurable energy density at
Earth (for an average modulation parameter $M=750$~MeV, see (1) in \rref{pg}).
This implies that less than half of the energy flux can be registered directly
at Earth.

In this article we will give an overview on recent experimental results and
their implications on the contemporary understanding of the origin of galactic
cosmic rays. As space is limited here, the reader may also consider further
recent reviews by the author \cite{pg,origin,cospar06,vulcano,ecrsreview}.

Progress in the understanding of the origin of galactic cosmic rays emerged 
mainly from observations in three complementary disciplines.
The direct measurement of cosmic-ray particles above the atmosphere in outer
space and on stratospheric balloons (\sref{direct}). 
At energies exceeding $10^{15}$~eV the steeply falling cosmic-ray energy
spectrum requires experiments with large detection areas exposed for long
times, at present, only realized in ground based installations
(\sref{indirect}). With such detectors extensive air showers are detected,
which originate in interactions of high-energy particles in the atmosphere
(\sref{eas}). 
And, finally, the observation of TeV $\gamma$-rays (\sref{gamma}).

\Section{Direct measurements} \label{direct}

Detectors above the atmosphere allow a precise determination of the composition
of cosmic rays and the measurement of energy spectra for individual elements,
presently, up to about $10^{14}$~eV. These findings are very important to
understand the propagation of the particles in the Galaxy.

In the 100~MeV energy domain detailed studies of the isotopic composition of
cosmic rays and the comparison to the local galactic abundance indicates that
cosmic rays are accelerated out of a sample of well mixed interstellar matter
\cite{cris-abundance}.  The propagation time of the particles in the Galaxy has
been determined to be $\tau_{esc}=(15.6\pm1.6)\cdot10^6$~yr
\cite{cris-time,garciamunoz75,garciamunoz88}.  At GeV energies the elemental
composition has been studied in detail, finding all elements of the periodic
table in cosmic rays, see e.g. \cite{cospar06}.  Some elements, as e.g. boron
are produced during the propagation in the Galaxy through spallation of heavier
elements such as carbon.  Investigations of the abundance ratios of different
elements (like the boron-to-carbon ratio) yield the column density of matter
traversed by the particles in the Galaxy \cite{cris-time,garciamunoz}.

\begin{figure}[t]
 \epsfig{file=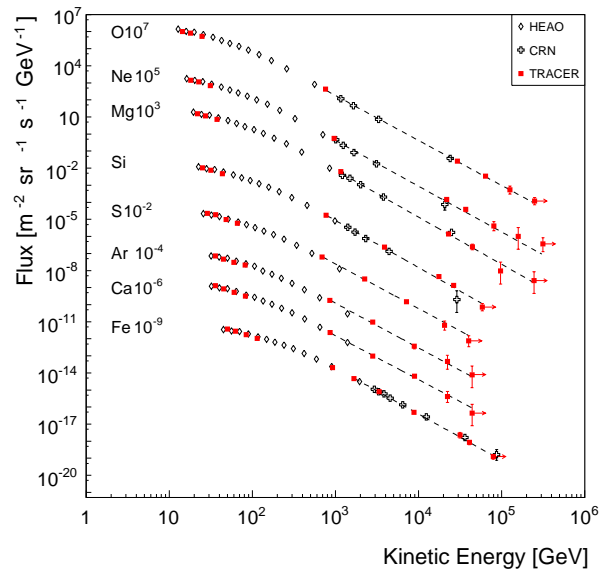,width=\columnwidth}
 \caption{Energy spectra for individual elements as measured by the TRACER,
	  HEAO, and CRN experiments \cite{boylemerida}.}
 \label{tracer}	  
\end{figure}

At present, several groups try to extend the direct measurements to higher
energies towards the knee (at about $4.5\cdot 10^{15}$~eV) in the all-particle
spectrum, benefiting from long-duration balloon flights \cite{ldb}.  Among
these are the ATIC \cite{atic}, CREAM \cite{creamexp}, and TRACER
\cite{gahbauer} experiments.  The latter is presently with an aperture of
5~m$^2$\,sr the largest cosmic-ray detector for direct measurements.  It
comprises several electromagnetic detectors, namely an acrylic \Cerenkov
counter, a proportional tube array, and a transition radiation detector, thus
covering a large energy range from 10~GeV/n up to $10^5$~GeV/n with
single-element resolution.  Spectra for elements from oxygen to iron as
measured by TRACER during an Antarctic circumpolar flight in 2003 are compiled
in \fref{tracer} \cite{boylemerida}.  The data extend previous results from the
space experiments HEAO (satellite) and CRN (on the Space Shuttle) to higher
energies. Currently, the data of a flight from Sweden to Alaska in 2006 are
analyzed. TRACER has been flown with an upgraded electronics being able to
record elements from boron to iron. The new data are expected to extend
measurements of the boron-to-carbon ratio up to $10^5$~GeV/n, thus, providing,
unprecedented information of the propagation properties approaching the knee in
the energy spectrum.

\begin{figure}[t]
 \epsfig{file=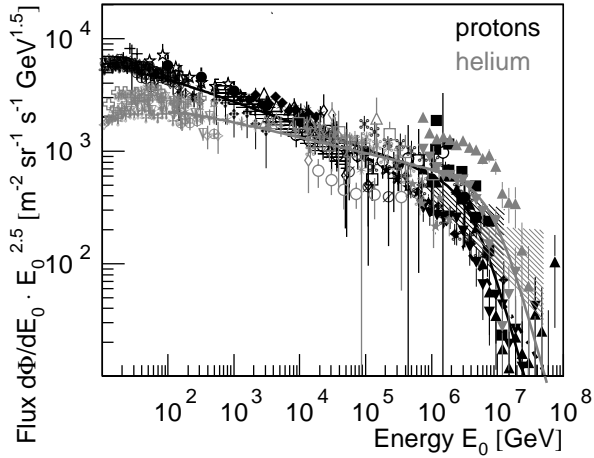,width=\columnwidth}
 \caption{Energy spectra for protons (black) and helium nuclei (grey).
	  For a list of experiments and references see \rref{cospar06}. 
	  The lines represent fits according to the \modell \cite{pg}.}
 \label{elementspek}	  
\end{figure}

Frequently, the question arises whether the energy spectra of protons and
helium have the same spectral index.  \Fref{elementspek} shows a compilation of
data, for details and references see \rref{cospar06}.  From the figure it
seems to be clear that the helium spectrum is indeed flatter as compared to
protons.  Due to spallation of nuclei during their propagation and the
dependence of the interaction cross section $\propto A^{2/3}$ one would expect
a slightly flatter spectrum for helium nuclei as compared to protons. Following
\cite{prop} the difference should be of the order of $\Delta\gamma\approx0.02$.
However, fits to the experimental data (lines in \fref{elementspek}) yield
$\gamma_p=-2.71 \pm0.02$ and $\gamma_{He}=-2.64\pm0.02$ \cite{pg}, resulting in
a difference $\Delta\gamma=0.07$.

\Section{Extensive air showers} \label{eas}

When high-energy cosmic-ray particles penetrate the Earths atmosphere they
interact and generate a cascade of secondary particles, the extensive air
showers. Two types of experiments may be distinguished to register air
showers: installations measuring the longitudinal development of showers (or
the depth of the shower maximum) in the atmosphere by observations of \Cerenkov
or fluorescence light and apparatus measuring the density (and energy) of
secondary particles (electrons, muons, hadrons) at ground level.

The shower energy is proportional to the total light collected or to the total
number of particles recorded at observation level.
More challenging is an estimation of the mass of the primary particle.
It is either derived by a measurement of the depth of the shower maximum \Xmax 
and the fact that the depth of the
shower maximum for a primary particle with mass $A$ relates to the depth of the
maximum for proton induced showers as 
\begin{equation} \label{xmaxeq}
 X_{max}^A=X_{max}^p-X_0\ln A, 
\end{equation}
where $X_0=36.7$~\gcm2 is the radiation length in air
\cite{matthewsheitler,jrherice06}.
Or, measuring the electron-to-muon ratio in showers. A Heitler model of
hadronic showers \cite{jrherice06} yields the relation
\begin{eqnarray} \label{emratioeq}
 \lg\left({N_e}/{N_\mu}\right)=C-0.065\ln A .
\end{eqnarray}

This illustrates the sensitivity of air shower experiments to $\ln A$.  To
measure the composition with a resolution of 1 unit in $\ln A$ the shower
maximum has to be measured to an accuracy of about 37~\gcm2 (see \eref{xmaxeq})
or the $N_e/N_\mu$ ratio has to be determined with an relative error around
16\% (see \eref{emratioeq}). Due to the large intrinsic fluctuations in air
showers, with existing experiments at most groups of elements can be
reconstructed with $\Delta\ln A\approx0.8-1$.

\begin{figure}[t]
 \epsfig{file=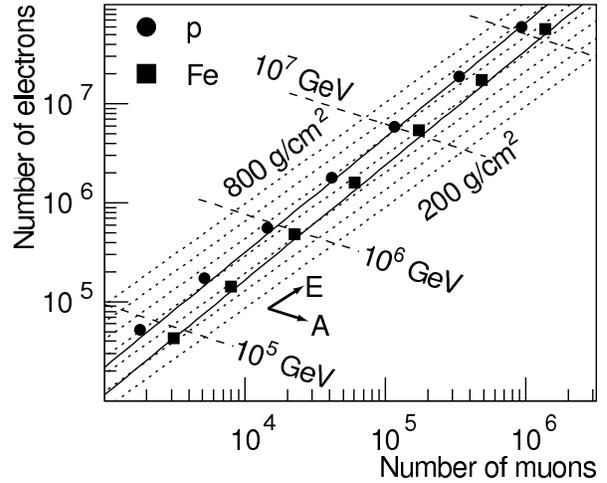,width=\columnwidth}
 \caption{Number of electrons vs. number of muons at shower maximum for fully
	  simulated showers (symbols). The lines represent predictions of a
	  Heitler model:
          solid -- constant mass for protons and iron nuclei \eref{masseq},
          dashed -- constant energy \eref{energyeq}, and
          dotted -- constant depth of the shower maximum \Xmax \eref{emxmaxeq}.}
 \label{emplane}
\end{figure}

The detection principle is illustrated in \fref{emplane}, depicting the number
of electrons as function of the number of muons at shower maximum. The symbols
represent fully simulated showers with discrete energies in steps of half a
decade, for details see \rref{jrherice06}. The lines represent predictions of a
Heitler model of hadronic air showers \cite{jrherice06}. The solid lines are
lines of constant mass
\begin{equation}\label{masseq}
 \left.N_e\right|_A=7.24\cdot A^{-0.16}N_\mu^{1.16}
\end{equation}
for primary protons and iron nuclei.
The dashed lines correspond to a constant energy, following
\begin{equation}\label{energyeq}
 \left.N_e\right|_{E_0} = 5.30\cdot10^7
            \left({E_0}/{\mbox{PeV}}\right)^{1.37}N_\mu^{-0.46}.
\end{equation}	    
The sets of lines define a parallelogram giving the axes for energy and mass in
the $N_e$-$N_\mu$ plane as indicated by the arrows.
The dotted lines represent a constant \Xmax, separated by 100~\gcm2 according
to
\begin{equation}\label{emxmaxeq}
 \left.\lg N_e\right|_{X_{max}}=
         \frac{X_{max}+287.9~\mbox{\gcm2}}{569.6~\mbox{\gcm2}}+1.02\,\lg N_\mu.
\end{equation} 

\begin{figure}[t]
 \epsfig{file=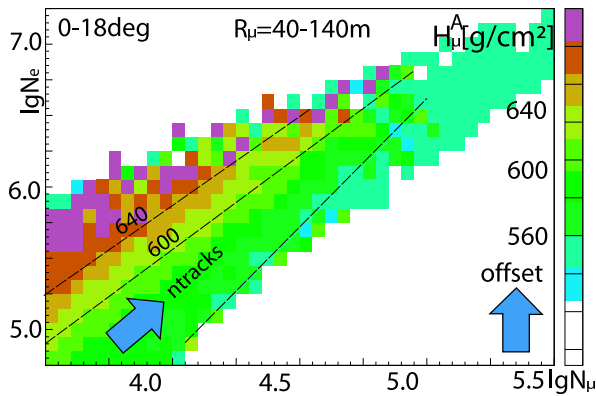,width=\columnwidth}
 \caption{Number of electrons vs. number of muons for showers measured with the
	  KASCADE experiment, the third dimension is the muon production hight
	  \cite{dollmerida}.}
 \label{mph}	  
\end{figure}

A similar plot is presented in \fref{mph}, showing the $N_e$-$N_\mu$ plane for
showers measured with the KASCADE experiment. The third dimension gives the
production hight of muons \cite{dollmerida}.
In the $N_e$-$N_\mu$ plane light primary elements are expected in the upper
left part of the figure. This is underlined by the measurements yielding in
this area deeply penetrating showers.
Attention should be payed when the two figures are compared: \Xmax for the
electromagnetic component (\fref{emplane}) is compared to \Xmax for the muonic
component ($E_\mu>0.8$~GeV, \fref{mph}).  \fref{emplane} represents $N_e$ and
$N_\mu$ at the shower maximum, \fref{mph} gives values measured at ground. For
muons the lateral distributions are integrated in the range interval from 40 to
200~m only in \fref{mph}. 

By detecting two quantities (e.g. $N_e$ and $N_\mu$ or $N_e$ and \Xmax), energy
and mass of the primary particles can be deduced. The arrival direction of
cosmic rays is determined by measurements of the arrival times of secondary
particles at the detectors.

\Section{Indirect measurements} \label{indirect}

Air shower experiments extend the findings of direct measurements to highest
energies. They provide information on the all-particle energy spectrum, the
(average) mass composition, and, recently, information on energy spectra for
elemental groups. These results contribute to the understanding of the origin
of structures in the all-particle spectrum, like the knee or the second knee.

The all-particle energy spectrum up to about $10^{18}$~eV is reasonably well
known. The spectra obtained by different experiments agree well in shape and
absolute flux taking a $\pm10\%$ uncertainty in the energy determination of the
individual experiments into account \cite{pg}.  The spectrum can be described
by a power law $dN/dE\propto E^\gamma$, with an spectral index changing at
the knee ($E_k=4.5$~PeV) from $\gamma=-2.7$ to $\gamma=-3.1$.

For the mean (logarithmic) mass of cosmic rays there seems to be a dependence
on the technique applied for the air shower observations \cite{wq}. Two classes
of experiments may be distinguished: detectors measuring the depth of the
shower maximum and experiments measuring (the lateral distribution of)
particles at ground level. This indicates inconsistencies in the interaction
models used to interpret the data.  For example, introducing lower inelastic
cross sections in the interaction model QGSJET~01 and slightly increasing the
elasticity, good agreement between the two classes of experiments can be
achieved \cite{wq}. Interpreting the data with this model yields a consistent
picture with a mean logarithmic mass smoothly continued from direct
measurements up to about $10^{15}$~eV, then exhibiting an increase up to
energies around $10^{17}$~eV.

\begin{figure}[t]
 \epsfig{file=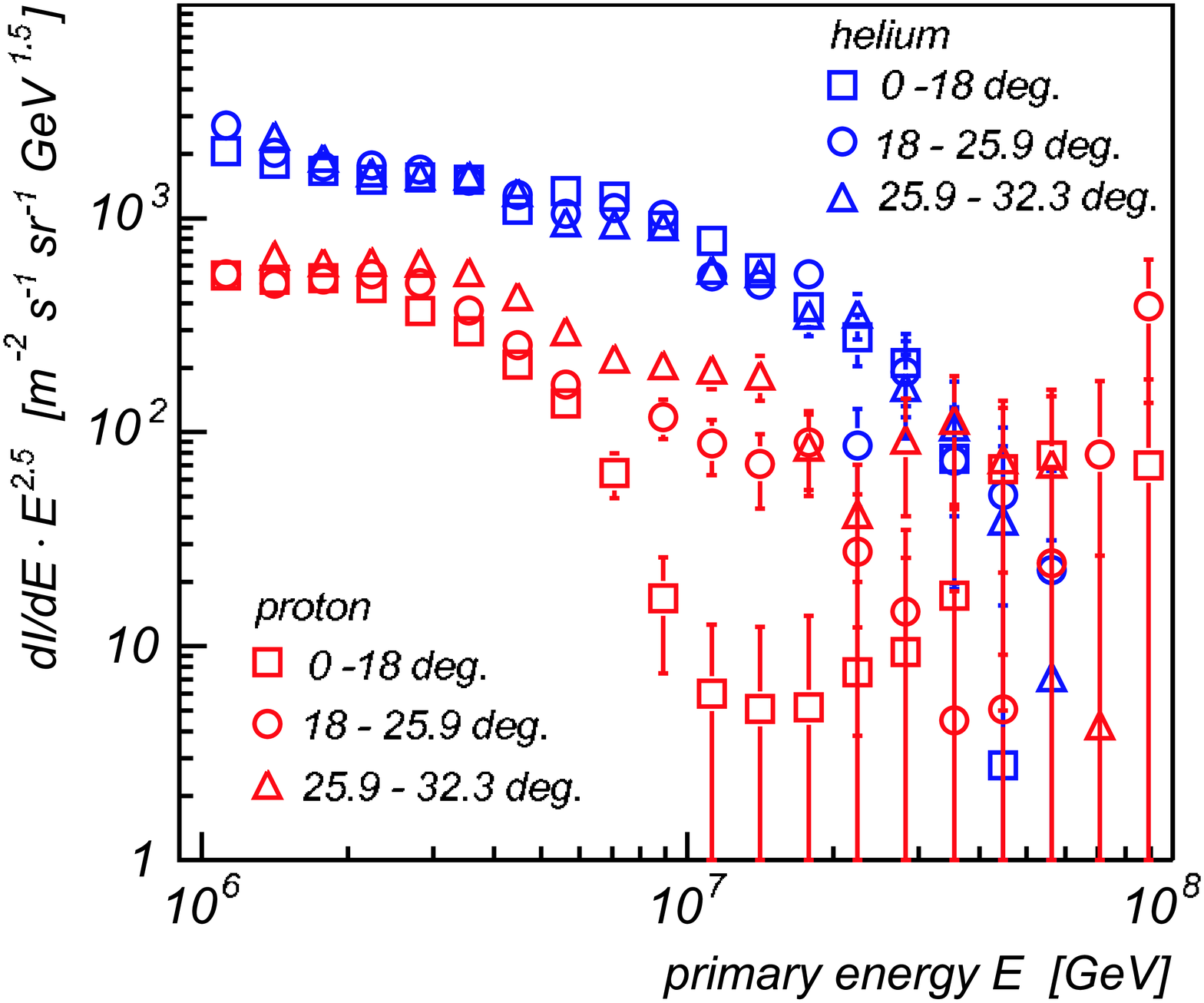,width=\columnwidth}
 \caption{Energy spectra for primary protons and helium nuclei, unfolded from
	  air showers registered by KASCADE within three zenith angle intervals
	  \cite{ulrichmerida}.}
 \label{finger}
\end{figure}

A big step forward in understanding the origin of the knee in the all-particle
energy spectrum are the results of the KASCADE experiment. The latter comprises
a $200\times200$~m$^2$ scintillator array to register the electromagnetic and
muonic component \cite{kascadenim}, a hadron calorimeter equipped with liquid
ionization chambers \cite{kalonim}, and a limited-streamer tube muon tracking
detector \cite{mtdnim}. The simultaneous detection of different shower
components allows detailed studies of hadronic interaction models
\cite{wwtestjpg,rissejpg,jenskrakow,jenspune,jensjpg,kascadewqpune,isvhecri02wwtest}.
The most consistent models are then applied to unfold energy spectra for five
groups of elements from the measured air shower data \cite{ulrichapp}.  The
energy spectra of the light elements (protons and helium) exhibit a depression
of the flux at high energies while the spectra of the heavier groups continue
to follow power laws to higher energies.  The results obtained using different
hadronic interaction models to interpret the data agree qualitatively. However,
there are differences in the absolute flux obtained, indicating inconsistencies
in the interaction models.  These results have recently been confirmed by an
analysis taking also inclined showers into account \cite{ulrichmerida}. As
example, the spectra obtained for primary protons and helium nuclei for three
different intervals of the cosmic-ray arrival direction are shown in
\fref{finger}. The different data sets are consistent with each other.

Similar energy spectra for groups of elements have been obtained by the EAS-TOP
\cite{eastopspec}, GRAPES-3 \cite{grapes05}, and Tibet \cite{tibetbdp}
experiments \cite{cospar06}.  Overall, a consistent 'standard picture' evolves
\cite{pg,cospar06}: the energy spectra as obtained by direct measurements can
be extrapolated to high energies, assuming power laws. The data are compatible
with cut-offs in the spectra proportional to their nuclear charge $Z$. The
all-particle spectrum and mean logarithmic mass thus obtained are compatible
with air shower data.  The combined energy spectra for groups of elements
obtained from direct and indirect measurements extend over seven orders of
magnitude in energy.  They are compatible with contemporary models of the
acceleration of cosmic rays in supernova remnants and models of the diffusive
propagation of the particles in the Galaxy \cite{cospar06}.

\begin{figure}[t]
 \epsfig{file=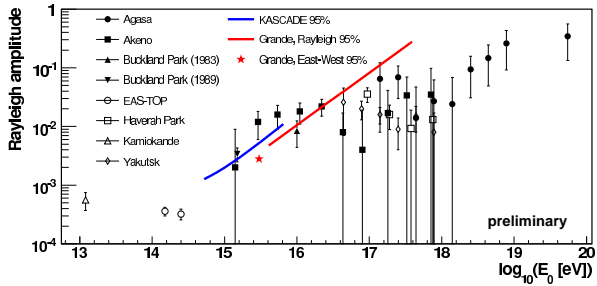,width=\columnwidth}
 \caption{Rayleigh amplitudes as function of energy as observed by different
          experiments \cite{overmerida}.}
 \label{aniso}
\end{figure}

Another important aspect is the anisotropy of the arrival direction of cosmic
rays, giving hints to understand the propagation of the particles in the
Galaxy. If the knee is caused by leakage of particles from the Galaxy, one
would expect less particles arriving from directions perpendicular to the
galactic disk.  Frequently, the Rayleigh formalism is applied to quantify the
observations \cite{kascade-aniso}.  Upper limits as obtained by KASCADE and
preliminary results obtained by the KASCADE-Grande experiment are compiled
together with results from many experiments in \fref{aniso} \cite{overmerida}.
An increase of the anisotropy amplitudes can be recognized. However, an
interpretation has to take into account that the available statistics decreases
roughly as $\propto E^{-3}$, therefore, an increase of the observed anisotropy
is expected, which is of the order of the observed increase.  The observed
anisotropy amplitude is compatible with a diffusion model of the particle
propagation \cite{kascade-aniso,maierflorenz,candiaaniso}.

\Section{TeV $\gamma$-ray astronomy} \label{gamma}

\begin{figure*}[t]\centering
 \epsfig{file=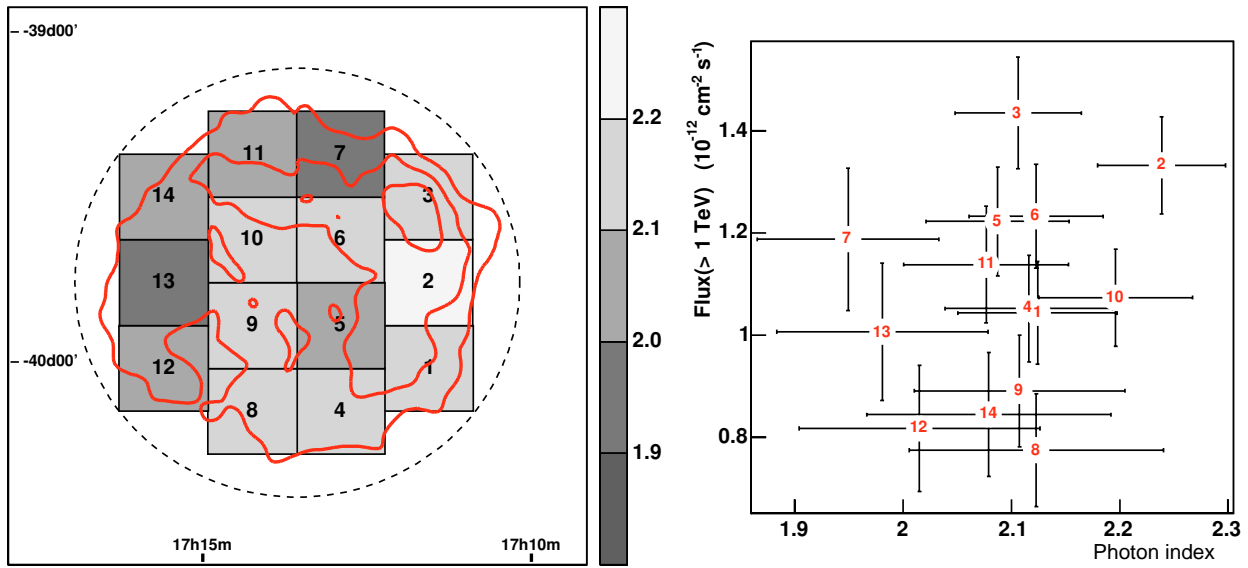,width=\textwidth}
 \caption{The SNR RX J1713.7-3946 as seen by the HESS experiment, spatially
	  resolved in TeV $\gamma$-ray light.  \LLeft: $\gamma$-ray excess
	  contours, superimposed are 14 fields for which spectra have been
	  obtained independently. The corresponding integral flux is displayed
	  as function of photon spectral index (\rright).  For details see
	  \cite{hessrxj1713}.}
 \label{snr}	  
\end{figure*}

Observation of TeV $\gamma$-rays with atmospheric \Cerenkov telescopes, like
the HESS \cite{hessexp}, MAGIC \cite{magic}, or VERITAS \cite{veritas}
experiments, give the most direct hints towards the acceleration sites of
cosmic rays.

The energy density of galactic cosmic rays amounts to about
$\rho_{E}\approx1$~eV/cm$^3$.  The power required to sustain a constant
cosmic-ray intensity can be estimated as $L_{cr}=\rho_{cr}
V/\tau_{esc}\approx10^{41}$~erg/s, where $\tau_{esc}$ is the residence time of
cosmic rays in a volume $V$ (the Galaxy, or the galactic halo).  With a rate of
about three supernovae per century in a typical Galaxy the energy required
could be provided by a small fraction ($\approx10\%$) of the kinetic energy
released in supernova explosions. This has been realized already by Baade and
Zwicky \cite{baadezwicky}.  The actual mechanism of acceleration remained
mysterious until Fermi \cite{fermi} proposed a process that involved
interaction of particles with large-scale magnetic fields in the Galaxy.
Eventually, this lead to the currently accepted model of cosmic-ray
acceleration by a first-order Fermi mechanism that operates in strong shock
fronts which are powered by the explosions and propagate from the supernova
remnant (SNR) into the interstellar medium \cite{axford,krymsky,bell,blanford}.

The theory of acceleration of (hadronic) cosmic rays at shock fronts in
supernova remnants is strongly supported by recent measurements of the HESS
experiment \cite{hesssnr,hessrxj1713}, observing TeV $\gamma$-rays from the
shell type supernova remnant RX J1713.7-3946, originating from a core collapse
supernova of type II/Ib. For the first time, a SNR could be spatially resolved
in $\gamma$-rays and spectra have been derived directly at a potential
cosmic-ray source as can be seen in \fref{snr}.  The figure shows on the
left-hand side excess contours of TeV $\gamma$-rays. The SNR has been divided
into 14 fields for which spectra have been obtained independently
\cite{hessrxj1713}. The corresponding integral flux values are plotted against
the spectral index on the right-hand panel.  The measurements yield a spectral
index $\gamma=-2.19\pm0.09\pm0.15$ for the observed $\gamma$-ray flux averaged
over the complete source. 

A model for the acceleration of hadronic particles in the SNR is supported by
measurements in various energy ranges from radio wavelengths to TeV
$\gamma$-rays \cite{voelkrxj1713}.  An important feature of the model is that
efficient production of nuclear cosmic rays leads to strong modifications of
the shock with large downstream magnetic fields ($B\approx100~\mu$G).  Due to
this field amplification the electrons are accelerated to lower maximum energy
and for the same radio/x-ray flux less electrons are needed.  Consequently, the
inverse Compton and Bremsstrahlung fluxes are relatively low only.  The model
predicts that the spectrum of decaying neutral pions, generated in interactions
of accelerated hadrons with material in the vicinity of the source, is clearly
dominant over electromagnetic emission generated by inverse Compton effect and
non-thermal Bremsstrahlung.  The measurements are compatible with a nonlinear
kinetic theory of cosmic-ray acceleration in supernova remnants and imply that
this supernova remnant is an effective source of nuclear cosmic rays, where
about 10\% of the mechanical explosion energy are converted into nuclear cosmic
rays \cite{voelkrxj1713,berezhkovoelksnr}.  

Further quantitative evidence for the acceleration of hadrons in supernova
remnants is provided by measurements of the HEGRA experiment \cite{hegra-casa}
of TeV $\gamma$-rays from the SNR Cassiopeia~A \cite{berezhko-casa} and by
measurements of the HESS experiment from the SNR "Vela Junior" (SNR RX
J0852.0-4622) \cite{voelksnr}.

For completeness, it should also be mentioned that the detection of
$\gamma$-rays with GeV energies contributed to the understanding of cosmic-ray
propagation.  The diffuse $\gamma$-ray background detected with the EGRET
satellite experiment \cite{egret} exhibits a structure in the GeV region, which
is interpreted as indication for the interaction of propagating cosmic rays
with interstellar matter \cite{strong-moskalenko}. This is the most direct
evidence for cosmic rays propagating in the galactic halo, well outside the
galactic disc.

\Section{Discussion and outlook} \label{outlook}
In the last decade our understanding of the origin of galactic cosmic rays has
been significantly improved by multidisciplinary efforts,
combining key observations
of the direct and indirect measurements of cosmic rays as well as the
detection of $\gamma$-rays.
The observations by the HESS $\gamma$-ray telescope give clear hints that
hadronic particles are accelerated in SNR. The data are compatible with a model
of first order Fermi acceleration at strong shock fronts. The particles
propagate in a diffusive process through the Galaxy. Parameters of the
propagation models have been constraint by direct measurements above the
atmosphere. The KASCADE experiment has shown that the energy spectra of the
light elements exhibit a cut-off structure, while the spectra of heavier
elemental groups follow power laws to higher energies. The observed spectra
seem to be compatible with the assumption of power laws and a cut-off energy
proportional to the nuclear charge. This implies that the knee in the
all-particle energy spectrum is caused by a cut-off of the light elements. The
shape of the all-particle spectrum at higher energies is then determined by the
subsequent cut-offs of all elemental species in cosmic rays. Most likely, the
astrophysical origin of the knee is a combination of the maximum energy reached
in the acceleration process and leakage from the Galaxy during propagation
\cite{prop}. More exotic ideas about the cause of the knee are most likely
excluded \cite{cospar06,ecrsreview}.

In conclusion this gives a qualitative 'standard picture' of the origin of
galactic cosmic rays.
However, several details remain unclear and a precise quantitative description
of all aspects of the acceleration and propagation mechanisms is still missing.
Among the open questions are: \\
It is not clear how to precisely match the spectral indices observed at Earth
to the spectra at the sources, being compatible with the TeV $\gamma$-ray
observations and the modification of the spectral slope during propagation.\\
A precise astrophysical interpretation of air shower data is at present limited
by the understanding of hadronic interactions in the atmosphere, thus the exact
shape of the energy spectra at their corresponding knees is unknown.\\
Contemporary assumptions on the parameters of cosmic-ray propagation models
yield anisotropies in the arrival directions not observed by air shower
experiments \cite{ptuskinaniso,prop}.

In the near future experiments like TRACER aim to reveal details of the
cosmic-ray propagation for energies approaching the knee \cite{muellermerida}.
Experiments at the LHC probing the extreme forward direction of phase space
will improve the description of high-energy hadronic interactions
\cite{engelpylos}.
Also the exploration of the end of the galactic component and the transition to
extragalactic particles in the energy range from $10^{17}-10^{18}$~eV will be
of importance. Key experiments in this energy region are KASCADE-Grande (a
0.5~km$^2$ extension of the KASCADE experiment) \cite{grande}, Ice Cube/ Ice
Top (a 1~km$^2$ air shower experiment and neutrino telescope at the South Pole)
\cite{icecube,icetop}, and HEAT/AMIGA (a 25~km$^2$ extension of the Auger
experiment to lower energies) \cite{klagesmerida}.

Around $10^{18}$~eV two features appear in the all-particle energy spectrum.
The second knee at $E_{2nd}\approx400$~PeV$\approx92\times E_k$, where the
spectrum exhibits a steepening to $\gamma\approx-3.3$, and the ankle at about
4~EeV, above this energy the spectrum seems to flatten again to
$\gamma\approx-2.7$. The region around 4~EeV is sometimes also called "the dip"
in the spectrum.\\
A possible cause for the second knee is the end of the galactic component, when
all elements successively have reached their cut-off energies, the latter being
proportional to their nuclear charge \cite{pg,prop}. If one assumes that
ultra-heavy elements (heavier than iron) play an important role to understand
the second knee, the factor of 92 between the energies of the knee and the
second knee can be easily understood as the nuclear charge of the heaviest
elements in the periodic table.\\
The dip is proposed to be caused by electron-positron pair production of cosmic
rays on photons of the cosmic microwave background \cite{berezinskydip}.

The investigation of the transition region and a precise measurement of the
galactic all-particle spectrum is also important for an estimate of the energy
content of extragalactic cosmic rays.  For example, the extragalactic component
needed according to the \modell to sustain the observed all-particle flux at
highest energies has an energy density of
$\rho_E^{exg}=3.7\cdot10^{-7}$~eV/cm$^3$. 

A promising rediscovered technique for the exploration of cosmic rays from the
transition region to highest energies is the detection of radio signals from
air showers \cite{allanrev}. Most likely the emission mechanism is coherent
synchrotron radiation of electrons with energies around the critical energy
(85~MeV) deflected in the magnetic field of the Earth (geosynchrotron
radiation) \cite{huegefalcke}. The LOPES experiment, registering showers in
coincidence with the KASCADE-Grande experiment has demonstrated the feasibility
of this approach \cite{radionature}. Radio detection of air showers is also
pursued in the CODALEMA experiment \cite{codalema} and within the LOFAR radio
telescope \cite{lofar}.  Also first radio pulses from air showers have been
recorded with antennae set up at the southern site of the Pierre Auger
Observatory \cite{vdbergmerida}.


\end{document}